\documentclass[aip,jcp,reprint,amsmath,amssymb]{revtex4-1}
\usepackage{color}
\usepackage{graphicx}
\usepackage[normalem]{ulem}
\usepackage{comment}

\newif\ifHighlitedChanges
\def\ifHighlitedChanges{\iffalse}
\ifHighlitedChanges
  
  \def\STRIKE#1{{\color{red}\sout{#1}}}
\else
  
  \def\STRIKE#1{\relax}
\fi
\begin{document}

\bibliographystyle{apsrev}

%\title{Heterogeneous Electron Transfer at Bithermal Molecule-Metal Interfaces}
%\title{Bithermal Electron Transfer at Molecule-Metal Interfaces}
%\title{Electron transfer at bithermal molecule-metal interfaces}
\title{Electron transfer at thermally heterogeneous molecule-metal interfaces}
\author{Galen T. Craven}
%\email[]{gcraven@sas.upenn.edu}
\affiliation{Department of Chemistry, University of Pennsylvania, Philadelphia, PA  19104, USA} 
\author{Abraham Nitzan}
%\email[]{anitzan@sas.upenn.edu}
\affiliation{Department of Chemistry, University of Pennsylvania, Philadelphia, PA  19104, USA} 
\affiliation{School of Chemistry, Tel Aviv University, Tel Aviv 69978, Israel}

%%%%%%%%%%%%%%%%%%%%%%%%%%%%%%%%%%%%%%%%%%%%%%%%%%%%%%%%%%%%%%%%%%%%%%%%%%%%%%%%%%
%\keywords{nonequilibrium dynamics, electron transfer, reaction rates, Marcus theory, transition state theory}
%%%%%%%%%%%%%%%%%%%%%%%%%%%%%%%%%%%%%%%%%%%%%%%%%%%%%%%%%%%%%%%%%%%%%%%%%%%%%%%%%%

\begin{abstract}
The rate of electron transfer between a molecular species and a metal, 
each at a different local temperature, 
is examined theoretically
through implementation of a 
%augmented 
bithermal (characterized by two temperatures) Marcus formalism.
Expressions for the rate constant and the electronic contribution to a
%emergent 
heat transfer mechanism which is
induced by the temperature gradient between molecule and metal are constructed.
The system of coupled dynamical equations describing
the electronic and thermal currents are derived and examined over 
diverse ranges of reaction geometries and temperature gradients. 
It is shown that electron transfer across the molecule-metal interface is associated with heat transfer and that the electron exchange between metal and molecule makes a distinct contribution to the interfacial heat conduction even when the net electronic current vanishes. 
%It is also observed that linear variation of the temperature gradient results in nonlinear behavior in the electron transfer rate.

\vspace{0.22cm}
\noindent This article may be downloaded for personal use only. 
Any other use requires prior permission of the author and AIP Publishing. 
This article appeared in \textit{J. Chem. Phys.} 146, 092305 (2017) and may be found at http://scitation.aip.org/content/aip/journal/jcp/146/9/10.1063/1.4971293
\end{abstract}
   \maketitle
\section{Introduction}
Molecular electronics \cite{Ratner1974rectifier,Carroll2002,Nitzan2003electron,Bredas2007}
provide a general platform to realize atomic-scale electronic and energy conversion devices through the control of electric currents
and thermal currents at molecule-metal interfaces.
%These molecular-level device architectures are a focus in 
%molecular engineering due to the possibility
%of fabricating circuits at increasingly smaller sizes,  
%and that operate at ever higher efficiency with respect to
%conventional electronics.
%A principal pathway for charge transport 
%in molecular devices
%is due to electric currents generated by
%electron transfer (ET) reactions
%between redox centers that change oxidation state
%due to electron charge localization.
Electronic transport through molecular junctions \cite{Nitzan2003electron,Nitzan2007jpcm,Dubi2011} is a process in which electrons move through the molecular network while interacting with the underlying nuclear environment.\cite{Nitzan2007,Galperin2009,Nitzan2011jpcl,Nitzan2011prb,Horsfield2006,DAgosta2008,Asai2011,Asai2015} The latter process give rise to inelastic effects in molecular electronic transport that may lead to heating and structural instabilities. The two extreme limits of this motion are, on one hand, elastic (tunneling and resonance) transport through the molecular electronic manifold in the absence of appreciable interaction with the nuclear environment, and on the other, a sequence of hopping processes through one or more intermediate redox sites on which the electron can be transiently localized by distorting its local nuclear environment. 
%For the case of homogeneous ET between two molecular species in a solvent environment,
%the Marcus picture,\cite{Marcus1956,Marcus1964,Marcus1985,Marcus1993,Peters2015}
%which represents each redox molecular state and corresponding solvent environment 
%as an (an)harmonic energy surface in a collective variable nuclear reaction coordinate,
%is the most prevalent theoretical methodology used to describe the process.
%In heterogeneous ET reactions between a molecule and 
%metal,\cite{Marcus1965,Voth1998,Finklea2001,Nitzan2006chemical,Nitzan2011,Compton2012,Bazant2014}
%the transfer process can be modeled by representing the electrode through a continuum of energy states 
%from which the electron can be transferred.\cite{Marcus1965,Wolf2007,Nitzan2012,Bazant2014}

The interplay between electronic and nuclear motions in controlling charge and energy transport through molecular junctions has been an active area of research for some time. \cite{Lake1992,Nitzan2007jpcm,Thoss2011,Lu2015} 
Junction heating (and its impact on junction stability) and heat transport  \cite{Cahill2002,Cahill2003,Leitner2008,Leitner2013,
Leitner2015,Li2012,Dhar2008,Luo2013,Rubtsova2015acr,Rubtsova2015} is one focus of these studies. \cite{Chen2003,Pecchia2007,Nitzan2007,Dubi2011,Li2012nano,Lu2015}
Thermoelectric energy conversion has been another. \cite{Dubi2011,Zimbovskaya2016}
Nonlinear effects such as heat current
rectification \cite{Li2004,Segal2005jcp,Chang2006,Segal2008prl,Segal2009}
and negative thermal resistance have been demonstrated, \cite{Zhong2009,Ren2013,Ming2016}
and possible ways to control heat transport in molecular junctions
have been discussed \cite{Arrachea2014,Donadio2015}.
%is induced by application of a temperature gradient across the junction. \cite{Reddy2007,Dubi2011,Lee2014}
%The heat transfer due to vibrational motion
%The use of tailored stimuli to induce specific transport properties in junctions
%is a general strategy toward optimal control in molecular circuitry.
%Perhaps most significant is the potential application of such studies to nanoscale thermoelectric devices. 
In addition to these advancements in charge transfer reactions across molecular junctions,
emergent experimental and theoretical methods examining the possibility to control electron transfer (ET) in 
specific vibrational modes \cite{Delor2014,Vlcek2015,Bredas2015mode} have also been developed.

Most studies of electron-vibration interaction in molecular junctions use the elastic transport 
as a starting point and treat inelastic effects as perturbations.\cite{Ren2012,Walczak2007,Koch2014,Perroni2014,Zimbovskaya2014} 
In the opposite limit, which describes electronic transport in redox molecular junctions, \cite{Migliore2013} 
electron transport can be described by a sequence of Marcus-type\cite{Marcus1956,Marcus1964,Marcus1985,Marcus1993,Peters2015} 
ET processes between the metal and molecular sites, and among molecular sites.  
While nuclear motion and reorganization are at the core 
of this ET mechanism,
the effect of thermal gradients, more generally thermal inhomogeneity, 
is not usually addressed for such processes. 
Similarly, while the implication of electron transport across interfaces on heat conduction in such systems have been often discussed, \cite{Nitzan2007jpcm,Dubi2011} 
such considerations are not usually made in the hopping transport limit. 

Recently, we have evaluated the effect of temperature difference between donor and acceptor sites on the rates of ET between them, as well as the contribution of the interfacial electron exchange to the interfacial heat transport. \cite{craven16c} 
Electron transfer was found to induce heat transfer between the donor and acceptor sites, and the ET rate was found to depend on both temperatures. 
This analysis can be generalized to consider the effect of thermal inhomogeneity in complex multithermal molecular reaction networks. \cite{craven17b}
 
In this article, we analyze a similar situation for ET between molecule and metal,
and between two metal electrodes through a molecular bridge,
in an electrochemical junction, generalizing the Marcus theory of ET between a metal electrode and a redox species in the adjacent solution to the case where the temperatures in the metal and molecule environments are different. 
It is %obviously 
relevant to thermoelectric transport in the hopping limit of molecular conduction, where the electron hops between different locations assumed to be in their own thermal equilibrium at their local temperatures. While hopping conduction is often invoked to describe electronic transport, its  implications for thermoelectric junctions has not yet been addressed.
The theory presented here provides a first step in this direction by providing a framework for describing electron transport across thermal gradients.  At the same time it advances our previous work on bithermal ET \cite{craven16c} to include reactions 
at thermally heterogeneous electrode interfaces, allowing implementation of the results in the design of general molecular-scale electronic components such as molecular wires and junctions.

%The principal focus of this article is to derive thermoelectric properties of molecular junctions \cite{Nitzan2003electron,Reddy2007,Malen2009,Malen2010}
%in the inelastic limit of electron transport.
In Sec.~\ref{sec:ET} the bithermal ET rate between a molecule and metal 
is derived, and we show how alteration of the temperature gradient between redox molecule/metal combinations
affects the interfacial thermoelectric properties.
Section~\ref{sec:HT} contains a derivation of the interfacial heat current between molecule and metal.
In Sec.~\ref{sec:EC} we combine the thermoelectric properties derived in previous sections
in order to describe the electric current and Seebeck coefficient in 
a prototypical model of a single molecule junction between two metal electrodes which are held at different temperatures.

\section{\label{sec:ET}Bithermal Electron Transfer at a Molecule-Electrode Interface}

\subsection{Electron transfer rates}

We consider a two-state ($a$ and $b$) ET process between a molecular species and a metal electrode.
%In state $a$, the molecular species is in a reduced state $\text{S}$ and the metal contains $N-1$ electrons.
%In state $b$, the molecular species is in an oxidized state $\text{S}^+$ and there are $N$ electrons on the metal.
%In state $a$, which corresponds to electron localization on the molecule, the molecular species is in a reduced state $\text{S}$. 
%State $b$ corresponds to electron localization in the energetic continuum of the metal and the molecular species being in an oxidized state $\text{S}^+$. 
For specificity, 
%in which
state $a$ corresponds to the molecular species being in a reduced state $\text{S}$,
and state $b$ corresponds to the molecular species being in an oxidized state $\text{S}^+$.
%Upon insertion of the electron into the metal, the free energy of the metal increases by a factor $\mu$, which is the chemical potential.
%Both the molecule and metal are in contact with an independent heat bath, each at a different temperature,
%and thus the studied system is bithermal (characterized by two temperatures).
The metal is assumed to be in its own electrochemical and thermal equilibrium characterized by the electrochemical potential $\mu$ and temperature $T_\text{M}$. The electronic population on the molecule interacts with its own equilibrium thermal environment, taken to be at a different temperature $T_\text{S}$
which is the temperature of the nuclei in the molecular environment.
%The temperature of the bath in contact with the molecule species is $T_\text{S}$ and the temperature of the metal is $T_\text{M}$.
The corresponding inverse thermal energies are $\beta_\text{S} = 1/k_\text{B} T_\text{S}$ and $\beta_\text{M} = 1/k_\text{B} T_\text{M}$ 
where $k_\text{B}$ is Boltzmann's constant.
Upon insertion of the electron into the metal, the free energy of the metal increases by an amount $\mu$.

If nuclear relaxation effects are ignored, the ET rates can be written as \cite{Nitzan2006chemical}
\begin{equation}
\label{eq:ETrateabnonuc}
\begin{aligned}
k_{a \to b} &=  \int_{\mathbb{R}}\big(1- f(\beta_\text{M},\epsilon)\big) \Gamma (\epsilon)  \delta(\epsilon-\Delta E_{ab})\,d\epsilon \\[1ex]
            &=  \big(1- f(\beta_\text{M},\Delta E_{ab})\big) \Gamma (\Delta E_{ab}) , 
\end{aligned}
\end{equation}
for the molecule to metal electron insertion process,
and 
\begin{equation}
\label{eq:ETratebanonuc}
\begin{aligned}
k_{b \to a} &=  \int_{\mathbb{R}}f(\beta_\text{M},\epsilon) \Gamma (\epsilon) \delta(\Delta E_{ab}-\epsilon)\,d\epsilon \\[1ex]
& = f(\beta_\text{M},\Delta E_{ab}) \Gamma (\Delta E_{ab}),
\end{aligned}
\end{equation}
for metal to molecule electron extraction,
where 
$f(\beta_\text{M},\epsilon) = (\exp\left[\beta_\text{M}(\epsilon-\mu)\right]+1)^{-1}$ is the
Fermi-Dirac distribution
characterizing the (assumed free-electron) metal and
$\Delta E_{ab}  = E'_a-E'_b$, with $E'_m \in \left\{a,b\right\}$ being an electronic occupation energy.
The integration interval $\mathbb{R}$ denotes integration over the region $(-\infty,\infty)$.
The single electron density of states in the metal $\rho_\text{M}$ and the tunneling coupling for electron transfer between molecule and metal
$V_{a,b}$
are both functions of $\epsilon$, and
\begin{equation}
\Gamma(\epsilon) = \bigg(\frac{2 \pi}{ \hbar} |V_{a,b}|^2 \rho_\text{M}\bigg)_\epsilon.
\end{equation}

With the inclusion of nuclear relaxation effects
the description of heterogeneous ET is fundamentally different.
This process is described below by adopting the Marcus formalism in which the energy surface 
representing each state is parabolic in a collective reaction coordinate $x$ 
that characterizes the nuclear degrees of freedom of the molecular species and its solvent environment.
In state $a$, the underlying potential surface is
\begin{equation}
\label{eq:Ea}
	E_a(x) =  \frac{1}{2} k (x-\lambda_a)^2 + E'_a,
\end{equation}
and in state $b$, 
\begin{equation}
\label{eq:Eb}
	E_b(x) =  \frac{1}{2} k (x-\lambda_b)^2 + E'_b,
\end{equation}
where $\lambda_m : m \in \left\{a,b\right\} $ are shifts in
the configuration associated with the two redox molecular states. 
%Only the difference $\lambda_a-\lambda_b$ is relevant to the present consideration.
This general formalism allows the accommodation a multitude of reaction geometries
through variation of the occupation energies and force constants. \cite{Marcus1965,Wolf2007,Bazant2014,Corni2016}
The reorganization energy of the ET reaction, which is independent of reaction direction, is
\begin{equation}
E_\text{R} = \frac{1}{2} k (\lambda_a-\lambda_b)^2.
\end{equation}
It has been observed that in molecule-metal ET reactions, 
e.g., in transition metal complexes,
the energy surfaces of the oxidized and reduced species can have
different curvatures. \cite{Hupp1984,Compton2012,Compton2013,Bazant2014} 
We ignore these asymmetric effects but note that 
the general formalism developed here can be modified to satisfy these physical situations through alteration of the
underlying energy surfaces.

The Marcus formalism describes the inelastic 
limit of electron transport in which relaxation of the
nuclear environment to a transient distorted state induced by electron localization occurs on a faster timescale  
than the electronic transition rate between molecule and metal sites, which is characterized by $\Gamma$.
%This the solvent environment of the molecular species, and thus
The strength of interaction between an electron and the nuclear environment of the 
solvent is characterized by the reorganization energy. 
When $E_\text{R}=0$, the transport is elastic and 
the electrons do not interact with the nuclear
environment. In the opposite inelastic limit, 
the energetic contribution of the reorganization energy to the transfer rate
depends on its relative weight which is dependent on the thermal energy of the molecular environment $k_\text{B} T_\text{S}$.

The transition under consideration is between the $a$ and $b$ states of the  molecule/metal.
Transfer can occur at all positions of the collective nuclear coordinate $x$ weighted by their thermal probability and subjected to the energy conservation constraint 
\begin{equation}
g_\text{c}(x,\epsilon) = E_b(x)-E_a(x) +\epsilon =0,
\end{equation}
where $\epsilon$ is the energy of the electron inserted to the metal.
%Thus, for each energy $\epsilon$, there is a unique value of the reaction coordinate, 
%\begin{equation}
%x^\ddag(\epsilon) = \frac{\Delta E_{ab}+ \epsilon}{k(\lambda_a-\lambda_b)} + \frac{\lambda_a+\lambda_b}{2}
%\end{equation}
%where the ET event can take place.
The corresponding ET rates are: from molecule to metal ($a$ to $b$ transition),
\begin{equation}
\label{eq:moltometal}
\begin{aligned}
k_{a \to b} &=  \iint_{\mathbb{R}^2} \big[1-f(\beta_\text{M},\epsilon)\big]\Gamma(\epsilon) \frac{\exp\big[-\beta_\text{S} E_a^\ddag(x)\big]}{Z_a^\ddag} \\
& \qquad \times \left|\nabla g_\text{c}\right|  \delta\big(g_\text{c}(x,\epsilon)\big) \,dx\,d\epsilon \\[1ex]
&= \sqrt{\frac{\beta_\text{S}}{4 \pi E_\text{R} }} \int_{\mathbb{R}}  \big[1-f(\beta_\text{M},\epsilon)\big] \Gamma(\epsilon)  \\
%& \qquad \times \exp\left[-\beta_\text{S} \frac{(\Delta E_{ba}-e \eta +\epsilon+E_\text{R})^2}{4 E_\text{R}}\right]\,d\epsilon
& \qquad \times \exp\left[-\beta_\text{S} \frac{(-\Delta E_{ab} +\epsilon+E_\text{R})^2}{4 E_\text{R}}\right]\,d\epsilon,
\end{aligned}
\end{equation}
and from metal to molecule ($b$ to $a$ transition),
\begin{equation}
\label{eq:metaltomol}
\begin{aligned}
k_{b \to a} &=  \iint_{\mathbb{R}^2} f(\beta_\text{M},\epsilon) \Gamma(\epsilon)  \frac{\exp\big[-\beta_\text{S} E_b^\ddag(x)\big]}{Z_b^\ddag} \\
& \qquad \times \left|\nabla g_\text{c}\right|  \delta\big(g_\text{c}(x,\epsilon)\big) \,dx\,d\epsilon \\[1ex]
&= \sqrt{\frac{\beta_\text{S}}{4 \pi E_\text{R} }} \int_{\mathbb{R}}  f(\beta_\text{M},\epsilon) \Gamma(\epsilon)  \\
%& \qquad \times \exp\left[-\beta_\text{S} \frac{(-\Delta E_{ba}-e \eta -\epsilon-E_\text{R})^2}{4 E_\text{R}}\right]\,d\epsilon
& \qquad \times \exp\left[-\beta_\text{S} \frac{(\Delta E_{ab} -\epsilon+E_\text{R})^2}{4 E_\text{R}}\right]\,d\epsilon,
\end{aligned}
\end{equation}
where the factor
$\left|\nabla g_\text{c}\right| = \left|k(\lambda_a-\lambda_b)\right|$ is the derivative magnitude of $g_\text{c}$
that removes ambiguity in the $\delta$-function constraint.
The function
\begin{equation}
\label{eq:act}
E_m^\ddag(x) = E_m(x) - E'_m: m \in \left\{a,b\right\}
\end{equation}
is the energy above the corresponding minimum and
\begin{equation}
\label{eq:part}
Z_m^\ddag= \int_{\mathbb{R}} \exp\big[-\beta_\text{S} E_m^\ddag(x)\big] \,dx = \sqrt{\frac{2 \pi}{\beta_\text{S} k }}:m \in \left\{a,b\right\}
\end{equation}
is the configuration integral 
associated with the molecule/solvent motion which depends on the temperature of
the molecular environment.
%Note that in Eqs.(\ref{eq:moltometal}) and (\ref{eq:metaltomol}) the vibrational energy is weighted by the thermal energy of the molecular mode $\beta_\text{S}$
%while the FD distribution is evaluated at the thermal energy of the metal $\beta_\text{M}$.
In the standard single temperature case ($T_\text{M} = T_\text{S} = T$), the results in Eqs.~(\ref{eq:moltometal}) and (\ref{eq:metaltomol})
reduce to the traditional Marcus-Hush-Chidsey rate expressions for heterogeneous ET \cite{Marcus1965,Hush1968,Chidsey1991,Nitzan2006chemical,Nitzan2011,Compton2012}
(cf. Eqs.~(17.11) and (17.12) in Ref.~\citenum{Nitzan2006chemical}). 

%%%%%%%%%%%%%%%%%%%%%
\begin{figure}
\includegraphics[width = 8.5cm,clip]{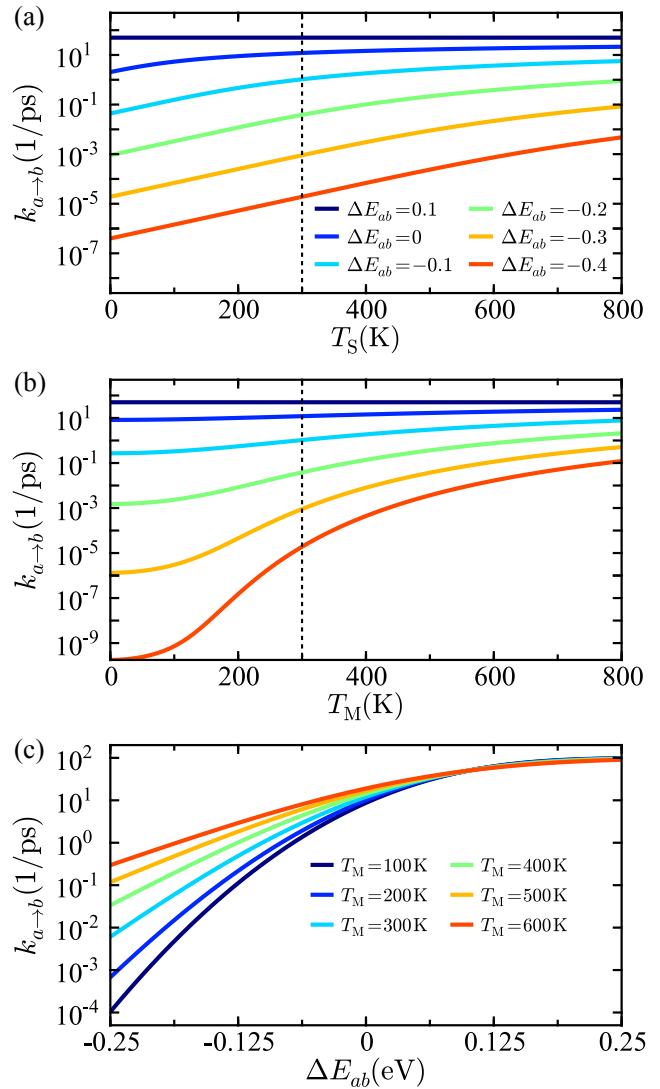}
\caption{\label{fig:rates}
Reaction rate $k_{a \to b}$ as a function of 
(a) $T_\text{S}$ with $T_\text{M} = 300\,\text{K}$ held constant 
and (b) $T_\text{M}$ with $T_\text{S} = 300\,\text{K}$ held constant.
Each curve is calculated for a different value of $\Delta E_{ab}$ shown
in the legend of (a) in units of $\text{eV}$.
The dashed vertical lines denote the unithermal ($T_\text{S} = T_\text{M}$) points.
(c) Reaction rate as a function of $\Delta E_{ab}$ for different values of $T_\text{M}$ shown
in the legend with $T_\text{S} = 300\,\text{K}$ held constant.
Parameters in all panels are  $\mu = 0$, $E_\text{R}  = 0.1\,\text{eV}$, and $\Gamma = 100\,\text{ps}^{-1}$. 
}
\end{figure}
%%%%%%%%%%%%%%%%%%%%%

Shown in Fig.~\ref{fig:rates} are the rates 
$k_{a \to b}$ computed for an example system 
over variation of the temperature of the metal $T_\text{M}$ and
temperature of the molecular environment $T_\text{S}$,
with all other parameters held constant.
As illustrated in Fig.~\ref{fig:rates}(a), varying $T_\text{S}$ with $T_\text{M}$ held constant
results in exponential dependence (linear on the semi-log scale) in the
low-temperature (relative to the temperature of the metal) regime of the molecular environment 
followed by crossover to nonlinear behavior in the logarithmic scale in the high-temperature regime.
The results of varying $T_\text{M}$ with $T_\text{S}$ held constant are 
shown in Fig.~\ref{fig:rates}(b).

Comparing Figs.~\ref{fig:rates}(a) and \ref{fig:rates}(b) it can be observed that changing the temperature of the metal results 
in a different functional form than variation of the 
temperature of the molecular environment (solvent).
In this case, the rate constant can be altered over orders of magnitude through relatively small variation of
the metal temperature.
This effect is particularly prominent for larger reaction free energies.
Examining the functional form in Eq.~(\ref{eq:moltometal}) and the corresponding results in Figs.~\ref{fig:rates}(a) and \ref{fig:rates}(b),
it can be seen that the reaction rate does not depend only on the magnitude of the temperature difference between molecule and metal,
but instead is a function of the specific temperature values.
%and the weight of the system energetics with respect to the thermal energies of each environment.

In Fig.~\ref{fig:rates}(c), the reaction rate is plotted over variation of $\Delta E_{ab}$ for
$T_\text{M}<T_\text{S}$, $T_\text{M}=T_\text{S}$, and $T_\text{M}>T_\text{S}$.
For $\Delta E_{ab}<E_\text{R}+\mu$, increasing the temperature of the metal results in an increase
in the reaction rate, which is the expected result.
This dependence changes at the point $\Delta E_{ab}=E_\text{R}+\mu$,
where $k_{a \to b}$
becomes independent of $T_\text{M}$. A reaction-rate turnover occurs for $\Delta E_{ab}>E_\text{R}+\mu$
in which the rate slightly increases with decreasing metal temperature.
Thus, in this limit, although this effect is small, the rate constants for systems of lower metal temperatures are larger than
that of systems with higher metal temperatures. 
Note that this is not the standard Marcus inverted regime (which is in fact absent in molecule-metal electron transfer),\cite{Migliore2012}
and it is unique to bithermal ET reactions
because the turnover occurs with respect to variation of the temperature of the metal, not variation of the free energy of the reaction.
In the limit
$\Delta E_{ab} \to \infty$, the reaction rate approaches an asymptotic value that does not depend on the temperature of the metal.

To explain the turnover behavior in the reaction rate with respect to variation in the metal temperature,  
consider the two oxidation states of the molecule: $\text{S}$ (electronic state $a$) and $\text{S}^+$ (electronic state $b$)
and the energy difference between the Marcus parabolas describing them
\begin{equation}
\begin{aligned}
\label{eq:Eabx}
E_{ab}(x) &= E_a (x)-E_b(x) \\[1ex]
&= k(\lambda_b-\lambda_a)x+\frac{1}{2} k \lambda_a^2-\frac{1}{2} k \lambda_b^2+\Delta E_{ab},
\end{aligned}
\end{equation}
which is linear in $x$ and $\Delta E_{ab}$. \cite{Migliore2012}
The energy differences at the two stable nuclear configurations of the system are 
$E_{ab}(\lambda_a)$ and $E_{ab}(\lambda_b)$.
%The rate $k_{a \to b}$ of the molecule changing from $\text{S}$ to $\text{S}^+$
%energy $E_{ab}(\lambda_a)$, 
In the regime $\Delta E_{ab}<E_\text{R}+\mu$, the transfer of an electron
from $E_{ab}(\lambda_a)$ into the metal
is energetically unfavorable and
increasing the metal temperature results in an increase in 
vacancies probabilities of the metal at energy levels below the Fermi level and about $E_{ab}(\lambda_a)$. 
This increases the probability for transfer into the metal,
and hence in this regime we observe the expected behavior that the reaction rate
increases with increasing metal temperature.
After the turnover point,
$\Delta E_{ab}>E_\text{R}+\mu$, and
electron transfer from level $E_{ab}(\lambda_a)$ into the metal is an energetically favorable transition.
Increasing the metal temperature decreases the number of vacant electronic states in the metal above energy $\mu$
and about $E_{ab}(\lambda_a)$,
which results in a decrease in the reaction rate.

The occupation probabilities for each state ($\mathcal{P}_a$ and $\mathcal{P}_b$) obey the kinetic equations
\begin{equation}
\begin{aligned}
\label{eq:occ}
\dot{\mathcal{P}}_a &= - k_{a \to b} \mathcal{P}_a+ k_{b \to a}\mathcal{P}_b, \\[1ex]
\dot{\mathcal{P}}_b &= - k_{b \to a} \mathcal{P}_b+ k_{a \to b}\mathcal{P}_a.
\end{aligned}
\end{equation}
%where $\dot{\mathcal{P}}_m :m \in \left\{a,b\right\}$ is proportional to the electric current.
At steady-state (ss), $\dot{\mathcal{P}}_a=0$ and $\dot{\mathcal{P}}_b=0$, and 
in this limit
%k_{a \to b} \mathcal{P}^{(\text{ss})}_a = k_{b \to a}\mathcal{P}^{(\text{ss})}_b$,
%which implies that the ratio of occupancy probabilities is,
\begin{equation}
K = \frac{\mathcal{P}^{(\text{ss})}_b}{\mathcal{P}^{(\text{ss})}_a} = \frac{k_{a \to b}}{k_{b \to a}}.
\end{equation}
In the absence of nuclear motion, 
%\begin{equation}
%K = \frac{1-f(\beta_\text{M},\Delta E_{ab})}{f(\beta_\text{M},\Delta E_{ab})}.
%\end{equation}
$K = \exp[-\beta_\text{M}(\mu-\Delta E_{ab})]$ is simply a ratio of Fermi distributions.
With the inclusion of nuclear effects from the solvent environment,
$K$ will depend on system parameters 
associated with the nuclear motion ($T_\text{S}$ and $E_\text{R}$).
The probability for the electron to occupy the molecule species is
\begin{equation}
\mathcal{P}^{(\text{ss})}_a = 1-\mathcal{P}^{(\text{ss})}_b = \frac{k_{b \to a}}{k_{a \to b}+k_{b \to a}}.  
\end{equation}

%%%%%%%%%%%%%%%%%%%%%
\begin{figure}
\includegraphics[width = 8.5cm,clip]{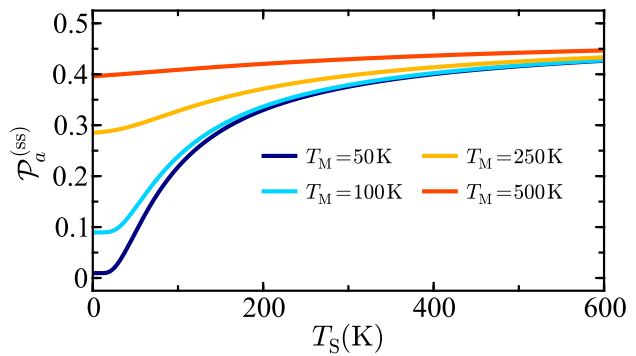}
\caption{\label{fig:pop}
Molecular occupation probability $\mathcal{P}^{(\text{ss})}_a$ 
%(solid)  and $\mathcal{P}^{(\text{ss})}_b$ (dashed) 
at steady-state as functions of $T_\text{S}$ for various values of $T_\text{M}$
shown in the legend.
Parameters are $\Delta E_{ab} = 0.01\,\text{eV}$, $\mu = 0$, $E_\text{R}  = 0.1\,\text{eV}$, and $\Gamma = 100\,\text{ps}^{-1}$.}
\end{figure}
%%%%%%%%%%%%%%%%%%%%%

%The steady-state occupation probabilities for a bithermal heterogeneous ET reaction 
%are shown in Fig.~\ref{fig:pop} over variation 
%of $T_\text{S}$ with $T_\text{M}$ held constant at different values.
%For small $T_\text{M}$, variation of $T_\text{S}$ can significantly alter the 
%probability of occupation on both the metal and the molecular species.
%In the high temperature limit of the molecular environment ($T_\text{S} \gg T_\text{M}$),
%$\mathcal{P}_a$ 
%approaches an asymptotic value
%and variation of the temperature difference results in linear behavior.
%These functional behaviors are in contradiction to the case of bithermal 
%molecule-to-molecule ET described in Ref.~\citenum{craven16c}
%where we have observed that 
%interchanging the temperatures between bithermal sites
%does not change the probability of occupation.
%Thus, we note that alteration of the probability of occupation
%by changing the temperature difference between sites in molecule-to-metal ET
%is caused by the temperature dependence of the occupation and vacancy distributions in the
%metal

The steady-state occupation probabilities for a bithermal heterogeneous ET reaction 
%with $\Delta E_{ab} = 0$ 
are shown in Fig.~\ref{fig:pop} over variation 
of $T_\text{S}$ with $T_\text{M}$ held constant at different values.
For high metal or molecule temperature ($k_\text{B} T_\text{S}$ or $k_\text{B} T_\text{M} \gg E_\text{R}, \Delta E_{ab}$),
 the molecular electronic population depends weakly on the temperature, however at low temperatures, this population is strongly affected by either $T_\text{S}$ 
or $T_\text{M}$. This stands in contrast to the corresponding effect in the case of molecule-to-molecule ET electron transfer examined in Ref.~\citenum{craven16c} where we have observed that when the two donor-acceptor sites are identical in energy and local vibrations, but differ in temperatures,
interchanging temperatures of the sites does not affect the
probability of occupation.
We next expand on the nature of this thermoelectric effect.

\subsection{Thermoelectric driving}

%%%%%%%%%%%%%%%%%%%%%
\begin{figure}
\includegraphics[width = 8.5cm,clip]{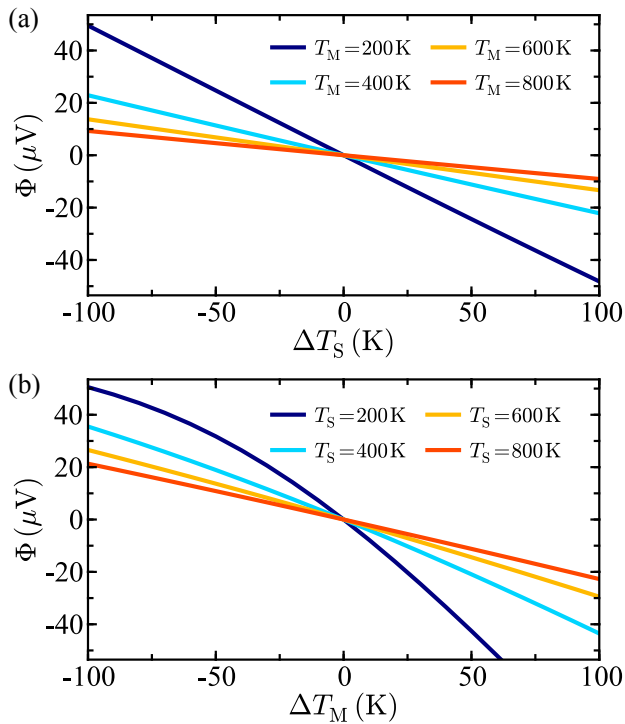}
\caption{\label{fig:thermo}
Electrostatic potential $\Phi$ to maintain zero current as function of $\Delta T$ for
(a) variation of $T_\text{S}$ with $T_\text{M}$ held constant 
and (b) variation of $T_\text{M}$ with $T_\text{S}$ held constant.
%The zero bias chemical potential $\mu_{I=0} = \mu$ is calculated at $\Delta T = 0$
%in all curves.
Parameters are $\Delta E_{ab} = -0.25\,\text{eV}$, $\mu = 0$, $E_\text{R}  = 0.1\,\text{eV}$, and $\Gamma = 100\,\text{ps}^{-1}$.}
\end{figure}
%%%%%%%%%%%%%%%%%%%%%

In analyzing electron transfer between two molecular sites of different temperatures, we have found that,
while the electron transfer rates are affected by both sites temperatures, 
there is no thermoelectric effect in the sense that temperature difference by itself does not drive electron transfer in a preferential direction. 
The reason for this behavior is that temperatures in this system are attributes of the nuclear environments, 
and in an otherwise symmetric system electron transfer in either direction is equally affected by both sites temperatures. 
The present situation is different, because one of the temperature considered (the metal's) reflects directly the occupation of electronic states. 
Thermoelectric driving is therefore expected. 
%To see its manifestation at the electode-metal interface we consider the dependence of the electrochemical potential   of the metal ($\Phi$ is the imposed electrostatic potential on the metal) as function of the temperature difference between metal and molecule under the condition that the electronic population on the molecule is constant (namely, that no current passes between molecule and metal).
To see its manifestation at the electode-metal interface, we consider the electrode potential $\Phi$ 
needed to maintain zero-current ($I=0$) as function of the temperature difference between metal and molecule.
% under the condition that the electronic population on the molecule is constant (namely, that no current passes between molecule and metal).

To calculate the zero-current bias between molecule and metal 
in the bithermal systems considered here,
the system is relaxed to 
%an electronic equilibrium state 
the zero current state
for particular values of 
$\Delta E_{ab}$, $\mu - e \Phi$, $T_\text{S}$, $T_\text{M}$, 
%and the resulting equilibrium constant is calculated. 
and the needed voltage $\Phi$ is calculated.
%We assume that the distance between redox sites is constant.
%The temperature of either the molecule or the metal is then varied
%the temperature difference $\Delta T_\text{S} = T_\text{S} - T_\text{M}$ and $\Delta T_\text{M} = T_\text{M} - T_\text{S}$ 
%and is then varied
%and the specific $\mu_{I=0} = \mu - e \Phi$ value needed to maintain equilibrium is calculated
%for the new set of temperatures.
This is performed for different molecule and metal temperatures, yielding $\Phi$ as function of these temperatures.
This gives a dependence of the resulting $\Phi$ as function of 
the temperature difference  
between molecule and metal;
a thermoelectric relation.

%Shown in Fig.~\ref{fig:thermo} are the electrostatic potentials needed to maintain a specific equilibrium state 
%for bithermal ET between metal and molecule as a function of the temperature difference.
%The resulting electrode potential PHI needed to maintain zero current between molecule and metal is shown in Fig. 3.
The resulting electrode potential $\Phi$ needed to maintain zero current between molecule and metal is shown in Fig.~\ref{fig:thermo}.
In Fig.~\ref{fig:thermo}(a) the molecular temperature is varied while the metal temperature is held constant.
In this case, the resulting $\Phi$ is linear in the temperature difference $\Delta T_\text{S} = T_\text{S} - T_\text{M}$ over all temperature variations.
Figure~\ref{fig:thermo}(b) illustrates the thermoelectric properties of the bithermal ET reaction over variation of 
the metal temperature, with the temperature of the molecular environment held constant.
Observe that over variation of $\Delta T_\text{M} = T_\text{M} - T_\text{S}$,
the resulting $\Phi$ is nonlinear, 
a contrast to the case of variation of $T_\text{S}$ with constant $T_\text{M}$ shown in Fig.~\ref{fig:thermo}(a).
Thus, and of significance, is the observation that the $\Phi$ needed to maintain $I=0$ does not depend on the absolute temperature difference, but is instead a quantity that 
varies independently with each temperature.
%This can be seen by comparing the corresponding curves for each $\mu_0$ value in Figs.~\ref{fig:thermo}(a) and \ref{fig:thermo}(b).
Note that the slopes of these curves are directly related 
to the Seebeck coefficient for the system. \cite{Reddy2007,Galperin2008,Ke2009,Liu2009,Sadat2010,Dubi2011,Tan2011,Kim2014,Lee2014,Koch2004,Koch2014,Simine2015,Zimbovskaya2016}
%In Fig.~\ref{fig:thermo}(a), over variation of $T_\text{S}$, the slopes are linear implying that the Seebeck coefficient is constant.
%Compare this with the case shown in Fig.~\ref{fig:thermo}(b) where variation of $T_\text{M}$ leads to 
%nonlinear curves 
%and thus the Seebeck coefficient is not constant.

\section{\label{sec:HT}Heat Current}
In bithermal heterogeneous ET reactions the temperature gradient of the system
can induce an interfacial heat current $\dot{\mathcal{Q}}$
between molecular environment and metal.
To derive this heat current,
consider the occupancy probability $\mathcal{P}_m$ that the system is in electronic state $m \in \left\{a,b\right\}$
and the conditional probability that the nuclear environment is in
a specific configuration $x$ given that the system is in state $m$:
\begin{equation}
P(x|m) = \frac{\exp\big[-\beta_\text{S} E_m^\ddag(x)\big]}{Z_m^\ddag} : m \in \left\{a,b\right\},
\end{equation}
where $E_m^\ddag(x)$ and $Z_m^\ddag$ are given by Eq.~(\ref{eq:act}) and Eq.~(\ref{eq:part}), respectively. 
We denote the joint probability distribution of these two independent events as
\begin{equation}
P(x,m) = P(x|m) \, \mathcal{P}_m : m \in \left\{a,b\right\}.
\end{equation}
The energy difference between surfaces describing the two electronic states is $E_{ab}(x)$
which by conservation of energy is the energy at which the electron enters/exits the metal during the ET process
at a particular configuration $x$.
%For the $a \to b$ transition, 
After the electron is transferred from molecule to metal ($a \to b$)
it equilibrates in the electronic manifold of the metal
depositing the amount $E_{ab}(x)-\mu$ of heat in the metal.
%and energy increases by a factor $\mu$. 
%Thus, the heat transferred during this process for a specific configuration $x$ is $E_{ab}(x)-\mu$.
%Correspondingly, 
Similarly,
for the metal to molecule  ($b \to a$) transition the heat transfer is $\mu - E_{ab}(x)$.
The contribution to the heat current from a particular nuclear configuration
will depend on the distribution $P(x,m)$,
the occupancy/vacancy probability of the metal at energy $E_{ab}(x)$,
which is given by the Fermi distribution,
and the ET rate at energy $E_{ab}(x)$.
The total heat current can be expressed as a product of these factors,
taken as a sum over all configurations
and over all possible state transitions.

For the $a \to b$ transition the heat current of the metal is
\begin{equation}
\begin{aligned}
\dot{\mathcal{Q}}^{(a \to b)}_\text{M} &= \int_{\mathbb{R}}  \big[1-f\big(\beta_\text{M}, E_{ab}(x)\big)\big] \Gamma \big(E_{ab}(x)\big) \\
& \quad \times \big( E_{ab}(x)-\mu \big)P(x,a)\,dx,
\end{aligned}
\end{equation}
and for the $b \to a$ transition
\begin{equation}
\begin{aligned}
\dot{\mathcal{Q}}^{(b \to a)}_\text{M} &= \int_{\mathbb{R}}  f\big(\beta_\text{M}, E_{ab}(x)\big) \Gamma \big(E_{ab}(x)\big) \\
& \quad \times \big(\mu -  E_{ab}(x)\big)P(x,b)\,dx.
\end{aligned}
\end{equation}
At steady state, $\mathcal{P}_m = \mathcal{P}^{(\text{ss})}_m$, and the number of $a \to b$ and $b \to a$ events per unit time are the same.
The net heat transfer for a pair of such transitions, $a \to b \to a$, is
%The total heat current is the sum of the two contributions:
\begin{equation}
\dot{\mathcal{Q}}_\text{M} = \dot{\mathcal{Q}}^{(a \to b)}_\text{M} + \dot{\mathcal{Q}}^{(b \to a)}_\text{M} = -\dot{\mathcal{Q}}_\text{S},
\end{equation}
%At the zero current steady-state $\mathcal{P}_m = \mathcal{P}^{(\text{ss})}_m$, 
%and in this limit 
%the heat currents can be written as
where
\begin{equation}
\begin{aligned}
\label{eq:heatcurrentAbe}
\dot{\mathcal{Q}}_\text{M} 
&=   \int_{\mathbb{R}} \mathcal{P}^{(\text{ss})}_a \big[1-f\big(\beta_\text{M}, E_{ab}(x)\big)\big] \Gamma \big(E_{ab}(x)\big) \\
& \quad \times \big( E_{ab}(x)-\mu \big) \frac{\exp\big[-\beta_\text{S} E_a^\ddag(x)\big]}{Z_a^\ddag}\,dx\\
&\quad +\int_{\mathbb{R}} \mathcal{P}^{(\text{ss})}_b  f\big(\beta_\text{M},E_{ab}(x)\big)\Gamma \big(E_{ab}(x)\big) \\
& \quad \times \big(\mu - E_{ab}(x)\big) \frac{\exp\big[-\beta_\text{S} E_b^\ddag(x)\big]}{Z_b^\ddag}\,dx. \\ 
%&= -\dot{\mathcal{Q}}_\text{S}.
\end{aligned}
\end{equation}
The relation $\dot{\mathcal{Q}}_\text{M}+\dot{\mathcal{Q}}_\text{S}=0$
(which is conservation of energy)
is shown explicitly in Appendix~\ref{sec:appendix}.

%with $E_{ab}(x) = E_a (x)-E_b(x)$.

%%%%%%%%%%%%%%%%%%%%%
\begin{figure}
\includegraphics[width = 8.5cm,clip]{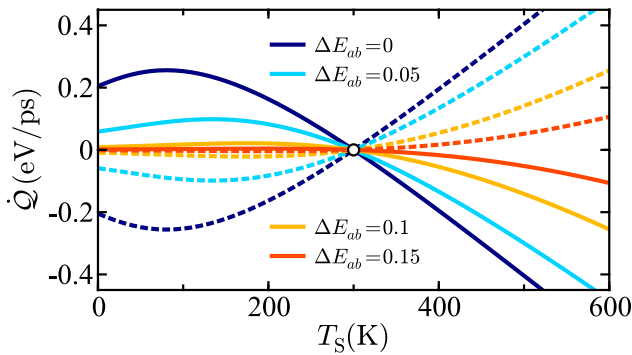}
\caption{\label{fig:current}
Heat current of the molecular environment  $\dot{\mathcal{Q}}_\text{S}$ (solid) 
and the metal $\dot{\mathcal{Q}}_\text{M}$ (dashed)
at steady-state as functions of $T_\text{S}$ with $T_\text{M} = 300\,\text{K}$ held constant.
Curves are shown for various values of $\Delta E_{ab}$ with colors corresponding to values shown in the legend in units of $\text{eV}$.
The circular marker denotes the unithermal point where $T_\text{M} = T_\text{S}$.
Parameters are $\mu = 0$, $E_\text{R}  = 0.1\,\text{eV}$, and $\Gamma = 100\,\text{ps}^{-1}$. }
\end{figure}
%%%%%%%%%%%%%%%%%%%%%

The steady-state heat currents induced by the temperature difference between molecule and metal 
are shown in Fig.~\ref{fig:current}
over variation of $T_\text{S}$ with $T_\text{M}$ held constant.
When the temperature of the molecular environment is less than the temperature of the metal, $T_\text{S}<T_\text{M}$, 
the heat current into the molecular environment is positive, $\dot{\mathcal{Q}}_\text{S}>0$, and the
heat current of the metal is negative,  $\dot{\mathcal{Q}}_\text{M} < 0$.
This is the expected result in which heat moves from the hot environment into the cold environment. 
At the unithermal point ($T_\text{M} = T_\text{S}$) the heat current vanishes.
%which illustrates that a net heat current between molecule and metal exists only in bithermal systems,
%and is absent at unithermal interfaces.
When $T_\text{S}>T_\text{M}$, 
the directionality of the heat current is reversed.
%and the
%current into the metal is positive
%and the current into the environment of the molecule is negative.
%Thus, in this limit,
%heat continues to move from hot to cold
%environments as required.
The same results for the heat currents can also be obtained using
expectation values for the amount of heat transferred by 
a single electron moving between molecule and metal.
See Appendix~\ref{sec:appendix} for details of this calculation.

\section{\label{sec:EC}Electric Current and Thermoelectricity}

\subsection{Electric current}

%The bithermal molecule-metal interface examined previously
%can be extended to model the electric current in a molecular junction which is characterized by 
%a temperature gradient between multiple electrodes.
%Here,
To see the implications of the above considerations on the transport properties of a redox molecular junction,
we consider a junction in which 
a molecular species with two electronic states ($a$ and $b$) is in contact with two metal leads.
The left ($\text{L}$) electrode has temperature
$T^\text{L}_\text{M}$, the right ($\text{R}$) electrode has temperature $T^\text{R}_\text{M}$, and
$\Delta T = T^\text{L}_\text{M} - T^\text{R}_\text{M}$.
The temperature of the molecular species is taken to be $T_\text{S} = (T^\text{L}_\text{M}+T^\text{R}_\text{M})/2$,
which is an assumption that arises from the postulates that 
the temperature gradient between the two metals is linear
and that the redox molecular site is seated a uniform distance from each electrode.
The chemical potentials of the metal electrodes are $\mu_\text{L} = \mu - e \Phi/2$ and $\mu_\text{R} = \mu + e \Phi/2$.

In this single-molecule two-electrode system, an electron whose charge is localized on the molecule can be transferred to either electrode,
and the forward and backward rate constants for these processes are given by evaluating Eqs.~(\ref{eq:moltometal}) and (\ref{eq:metaltomol})
at the corresponding temperatures and chemical potentials. 
For the left electrode $k^\text{L}_{a \to b}$ and $k^\text{L}_{b \to a}$ are evaluated at $T_\text{M}= T^\text{L}_\text{M}$,
and for the right electrode $k^\text{R}_{a \to b}$ and $k^\text{R}_{b \to a}$ are evaluated at $T_\text{M}= T^\text{R}_\text{M}$.
The kinetic equations describing the occupation probabilities of states $a$ and $b$ are
\begin{equation}
\begin{aligned}
\label{eq:occ2}
\dot{\mathcal{P}}_a &= - \left(k^\text{L}_{a \to b}+k^\text{R}_{a \to b}\right) \mathcal{P}_a+ \left(k^\text{L}_{b \to a}+k^\text{R}_{b \to a}\right)\mathcal{P}_b, \\[1ex]
\dot{\mathcal{P}}_b &= - \left(k^\text{L}_{b \to a}+k^\text{R}_{b \to a}\right) \mathcal{P}_b+ \left(k^\text{L}_{a \to b}+k^\text{R}_{a \to b}\right)\mathcal{P}_a.
\end{aligned}
\end{equation}
At steady state, the populations of each state can be expressed as
\begin{equation}
\mathcal{P}^{(\text{ss})}_a = 1-\mathcal{P}^{(\text{ss})}_b =  \frac{k^\text{L}_{b \to a}+k^\text{R}_{b \to a}}{k^\text{R}_{a \to b}+k^\text{R}_{b \to a}+k^\text{L}_{a \to b}+k^\text{L}_{b \to a}},  
\end{equation}
and the steady-state electronic current $I$ is \cite{Nitzan2011}
%and the current $I$  can be written in terms of the rate constants as
\begin{equation}
\label{eq:electriccurrent}
\frac{I}{e} = \frac{k^\text{L}_{a \to b}k^\text{R}_{b \to a}-k^\text{R}_{a \to b}k^\text{L}_{b \to a}}{k^\text{R}_{a \to b}+k^\text{R}_{b \to a}+k^\text{L}_{a \to b}+k^\text{L}_{b \to a}}.
\end{equation}

%%%%%%%%%%%%%%%%%%%%%
\begin{figure}
\includegraphics[width = 8.5cm,clip]{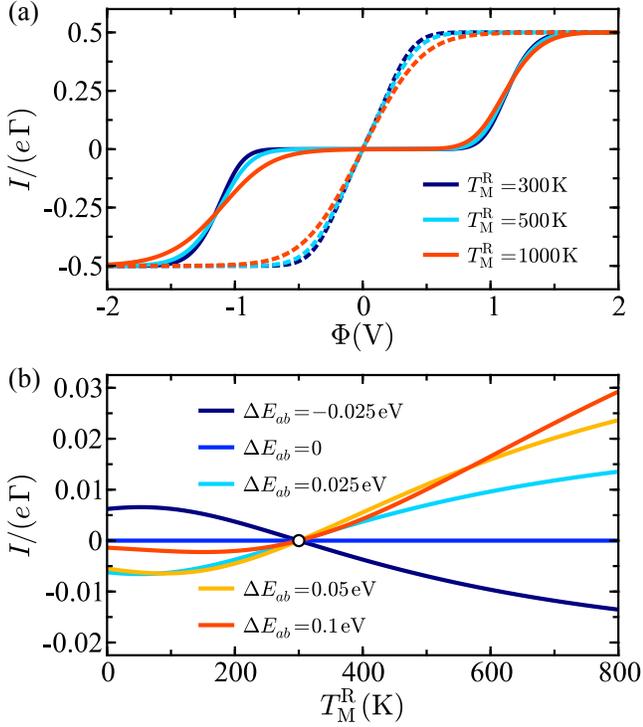}
\caption{\label{fig:electriccurrent}
Electric current $I$ as a function of 
(a) electrostatic potential $\Phi$ for varying $T^\text{R}_\text{M}$
with $\Delta E_{ab} = -0.5\,\text{eV}$ (solid) and $\Delta E_{ab} = 0$ (dashed),
and (b) right electrode temperature $T^\text{R}_\text{M}$ with $\Phi$=0 and various values of $\Delta E_{ab}$ shown in the legend.
The circular marker denotes the unithermal point where $T^\text{L}_\text{M} = T^\text{R}_\text{M}$.
Parameters are $T^\text{L}_\text{M} = 300\,\text{K}$, $\mu = 0$, $E_\text{R}  = 0.1\,\text{eV}$, and $\Gamma = 100\,\text{ps}^{-1}$.}
\end{figure}
%%%%%%%%%%%%%%%%%%%%%

%The electric current generated in the junction is shown in 
This current is shown in Fig.~\ref{fig:electriccurrent}
as function of different system parameters.
In Fig.~\ref{fig:electriccurrent}(a) it is 
shown as a function of $\Phi$
for various values of $T^\text{R}_\text{M}$ with $T^\text{L}_\text{M} = 300\,\text{K}$ held constant.
For $\Delta E_{ab} = 0$, the current is symmetric in the applied voltage $I(\Phi) = I(-\Phi)$,
which is an obvious consequence from the symmetry of the structure.
However, when $\Delta E_{ab} \neq 0$ and $T^\text{L}_\text{M} \neq T^\text{R}_\text{M}$, this symmetry is destroyed.
The reason for this is that the contribution to the current induced by the temperature difference depends on the sign of
$\Delta E_{ab}$ as explained below.
For $\Delta E_{ab} \neq 0$, $I(\Phi) \neq I(-\Phi)$),
illustrating that asymmetrical effects generated
in the junction due to the temperature gradient are dependent on the free energy difference
between electronic states
in the molecule.

%Shown in Fig.~\ref{fig:electriccurrent}(b) is the electronic current
%as a function of $T^\text{R}_\text{M}$ at zero bias ($\Phi = 0$) and varying values of $\Delta E_{ab}$.
%For $\Delta E_{ab}=0$, over all values of  $T^\text{R}_\text{M}$,
%there is no electric current, which illustrates that in systems without electronic driving, either from a potential difference across electrodes or from the 
%energy difference in molecular oxidation states, there is no net electronic flow, even when  $T^\text{L}_\text{M} \neq  T^\text{R}_\text{M}$.
%For $\Delta E_{ab} \neq 0$, electric currents are observed
%and general trends can be noted.
%When $T^\text{R}_\text{M}<T^\text{L}_\text{M}$
%the current flows from $\text{L} \to \text{R}$, 
%as expected due to the increased occupancy of energy levels above $\mu$ in 
%the left electrode due to the higher temperature, with respect to the colder right electrode.
%At the unithermal point where $T^\text{R}_\text{M} = T^\text{L}_\text{M}$
%the current stops.
%When the temperature of the right electrode is increased from the unithermal point ($T^\text{R}_\text{M}>T^\text{L}_\text{M}$),
%the current flows in the opposite direction, from $\text{R} \to \text{L}$.

Shown in Fig.~\ref{fig:electriccurrent}(b) 
is the electronic current as function of $T^\text{R}_\text{M}$, keeping $T^\text{L}_\text{M}$ constant, 
at zero bias ($\Phi=0$)
for different values of $\Delta E_{ab}$. 
To understand the observed behavior it should be noted that $E_{ab}(x)$, given by Eq.~(\ref{eq:Eabx}), 
corresponds in our model to the single electron energy (the occupation energy) associated with the molecule at nuclear configuration $x$, 
and its effect on electron transmission depends on the difference $E_{ab}(x)-\mu$. 
In the present model, where nuclear reorganization is represented by shifted harmonic surfaces, $\Delta E_{ab} =0$ corresponds
(for the present choice of $\mu=0$)
to the  case where $E_{ab}(\lambda_a)=-E_{ab}(\lambda_b)$, 
namely to the situation where the single electron ``molecular level'' at the equilibrium nuclear positions of the occupied state 
$\lambda_a$ and the unoccupied state $\lambda_b$ are symmetrically seated above and below the Fermi level. 
This implies that the electron and hole currents are equal in this situation which explains the vanishing of the net current seen in this case.  
For $\Delta E_{ab} \neq 0$, the direction of the thermoelectric current (hot to cold or vice versa) depends on the sign of $\Delta E_{ab}$ - an extension of the behavior known for electron or hole dominated currents in molecular thermoelectrics.

Nonmonotonic behavior in the electric current can also be observed in Fig.~\ref{fig:electriccurrent}(b) 
with respect to variation of the energy difference between electronic states in the molecular species.
In the low-temperature limit ($T^\text{R}_\text{M} \to 0$),
the magnitude of the current $|I|$ decreases with increasing $|\Delta E_{ab}|$.
In temperature regimes both above and below the unithermal point,
the electric current exhibits nonmonotonic trends 
in which the ordering of $|I|$ with respect to $|\Delta E_{ab}|$ 
is dependent on the specific value of $T^\text{R}_\text{M}$.
In the high-temperature limit ($T^\text{R}_\text{M} \to \infty$), 
increasing $|\Delta E_{ab}|$ results in an increased current magnitude.

\subsection{Seebeck coefficient}

The standard Seebeck coefficient $S$ measures the dependence of the voltage across the junction
on the temperature difference between the left and right electrodes, \cite{Galperin2008,Reddy2007,Tan2011} 
calculated about equilibrium under the condition of constant, namely zero, current:
\begin{equation}
\label{eq:Seebeck}
S = -\left(\frac{d \Phi}{d \Delta T}\right)_{\text{eq},I=0}.
\end{equation}
This is most easily evaluated using Eq.~(\ref{eq:electriccurrent})
and the identity
\begin{equation}
\label{eq:See2}
-\left(\frac{d\Phi}{d\Delta T}\right)_I 
= \left(\dfrac{\partial I}{\partial \Delta T} \right)_{\Phi} \Bigg/\left(\dfrac{\partial I}{\partial \Phi}\right)_{\Delta T},
\end{equation}
with all derivatives evaluated at $I = \Delta T = \Phi = 0$.
\begin{figure}
\includegraphics[width = 8.5cm,clip]{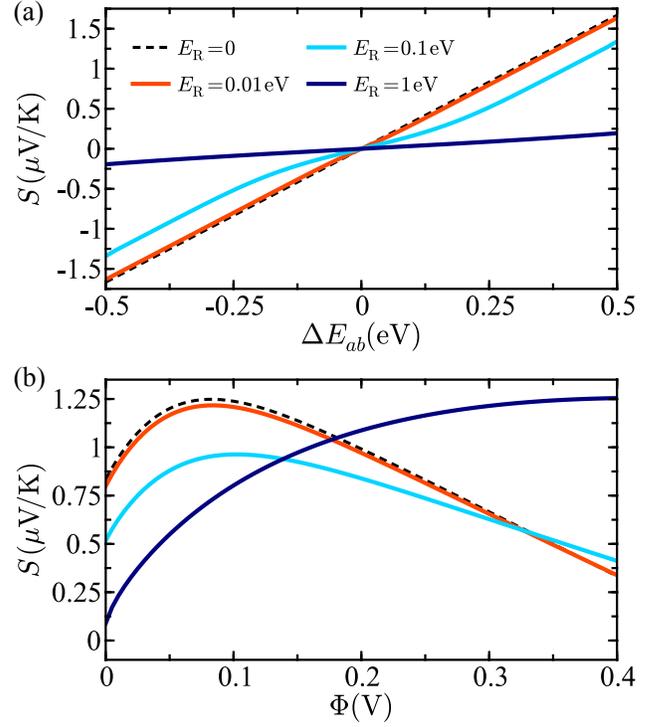}
\caption{\label{fig:Seebeck}
Seebeck coefficient $S$ as function of (a) $\Delta E_{ab}$ 
and (b) $\Phi$ with $\Delta E_{ab} = 0.25\,\text{eV}$, for
various reorganization energies.
%$E_\text{R}$ shown in the legend.
%with $E_\text{R} = 0$ shown as a dashed curve.
Parameters are $T^\text{L}_\text{M} = 300\,\text{K}$, $T^\text{R}_\text{M} = T^\text{L}_\text{M} - \Delta T$, $\mu = 0$, and $\Gamma = 100\,\text{ps}^{-1}$.\cite{note2}}
\end{figure}
%%%%%%%%%%%%%%%%%%%%%

The calculated standard ``equilibrium'' Seebeck coefficient is shown 
in Fig.~\ref{fig:Seebeck}(a)
as a function of $\Delta E_{ab}$.
In the limit $E_\text{R} = 0$, it is $(\Delta E_{ab} - \mu)/(e T_\text{M}^\text{L})$, 
which is easily obtained from Eqs.~(\ref{eq:ETrateabnonuc}) and (\ref{eq:ETratebanonuc}).
When $E_\text{R} \neq 0$, $S$ becomes smaller and is a slightly nonlinear function of $\Delta E_{ab}$
(note however that linearity is restored for large $\Delta E_{ab}$).
%Note that $S$ is approximately linear at large values of $|\Delta E_{ab}|$ for large $E_\text{R}$,
%hich agrees with the  $E_\text{R} = 0$ results
%The sign of $S$ depends on the sign of $\Delta E_{ab}$,
%and for $\Delta E_{ab}<0$, $S$ is negative, while, correspondingly,
%for $\Delta E_{ab}>0$, $S$$ is positive, as expected.
As expected, $S$ changes sign with $\Delta E_{ab}$ which measures
the position of the molecular ``single electron level'' relative to the metal chemical potential.
%Note that the Seebeck coefficient is 
%an odd function of energy difference between electronic states of the molecule,
%and thus $S(\Delta E_{ab}) = -S(-\Delta E_{ab})$ which is a consequence of symmetry.
As $\Delta E_{ab} \to 0$, the thermopower in the junction vanishes,
implying that without an energy gradient in the electronic states of the molecule, 
the electric current will vanish, regardless of the imposed temperature difference, which agrees with the results shown in Fig.~\ref{fig:electriccurrent}(b).
Increasing the reorganization energy $E_\text{R}$, 
which is a measure of the nuclear-electronic coupling,
results in smaller values of $S$, 
illustrating that stronger coupling
%inhibits electronic transport
leads to lower thermopower in the junction. 
%This is caused by timescale relaxations nuclear interactions inhibit the 
%transport in the junction 
%in the junction.
%while for $|\Delta E_{ab}| \sim E_\text{R}$
%nonlinear behavior is observed.

The Seebeck coefficient can be calculated outside of the linear $I \to 0$ limit using the relation \cite{Galperin2008}
\begin{equation}
\label{eq:Seebecknon}
S(I) = -\frac{\Phi(I)}{\Delta T(I)},
\end{equation}
where $\Delta T$ is the temperature difference that generates the same current at $\Phi = 0$, as
the $\Phi$ generates for $\Delta T = 0$.
Equation~(\ref{eq:Seebecknon}) is a generalization of the standard definition of the Seebeck coefficient as an attribute of the equilibrium junction to 
linear response about an arbitrary equilibrium point.
The protocol we apply to measure $S$ is to change $T^\text{R}_\text{M}$ while keeping $T^\text{L}_\text{M}$ constant
for $\Phi=0$, and to apply the bias symmetrically across the junction ($\mu_\text{L} = \mu - e \Phi/2$ and $\mu_\text{R} = \mu + e \Phi/2$)
for $\Delta T = 0$. \cite{note2}
Shown in  Fig.~\ref{fig:Seebeck}(b) is $S$ as a function of $\Phi(\Delta T = 0, I)$, which is the inverse function 
of $I(\Delta T = 0, \Phi)$, for different reorganization energies and constant $\Delta E_{ab}$.
At $\Phi=0$, the value of $S$ is the same as that shown in Fig.~\ref{fig:Seebeck}(a)
for the corresponding value of $\Delta E_{ab}$ and $E_\text{R}$.
As $\Phi$ is increased, the Seebeck coefficient increases nonlinearly,
which agrees with the behavior observed in Ref.~\citenum{Galperin2008} 
in molecular junctions at the inelastic limit of transport.
As $\Phi$ is increased further, a turnover is observed for small values of $E_\text{R}$ (weak electron-environment interaction), and 
$S$ begins to decrease. 
This is the same trend that has been observed previously in studies of 
transport in junctions in the weak electron-phonon coupling limit.\cite{Galperin2008}
%Sentence about  reorganization energy dependence.
%In the limit $\Phi \to \infty$, the thermopower in the junction vanishes ($S \to 0$).

\section{Conclusions}

A theory has been developed to describe the rate of 
electron transfer between a molecular species and a metal electrode, 
with each being at a different local temperature.
The rate constant for this process was found to be nonlinear in the temperature of each environment.
%and not a function simple fucntion of the temperature difference. 
We find that due the temperature gradient,
electron transfer between redox sites
carries heat between 
the metal and the thermal environment of the molecule,
and this contribution to the interfacial heat conduction has been characterized.
Analogous to previous results observed in bithermal molecule-to-molecule electron transfer reactions, \cite{craven16c}  
the electrothermal heat transfer does not vanish when the electric current between molecule and metal reaches a stationary state.

Thermoelectric effects induced by a temperature difference between heterogeneous
redox sites have also been investigated.
The findings presented here illustrate how electronic and thermal transport 
are related at the strong-coupling limit, and
how electrothermal transport and traditional thermoelectric effects can be induced.
Control of transport and amplification in thermal currents 
has potential applications in the development of novel energy conversion devices and molecular electronics.
Operational deficiencies in thermal logic gates and circuits with respect to their electronic analogs
occur due to timescale mismatches in phononic transport, which takes place on the time-scale of nuclear motion, and electronic transport.
%which occurs on the time-scale of electron motion.
The theory of electrothermal transport presented here can possibly provide rectification of this timescale problem in 
thermal circuits due to the described heat transfer mechanism occurring on the timescale of electron motion.
Further study and validation of this conjecture is required, and
in future work we will provide a rigorous comparison of the rates and timescales of phononic and electronic heat transport.

The presented results 
%which examine charge and heat transfer between a single molecule and both one and two electrodes,
provide a step toward the ability to completely model electron hopping in molecular junctions
in which complex molecular motifs are seated between two electrodes. 
In future work, we will also present a theory
for bithermal electron transfer at the weak-coupling limit in which
electron transmission between redox sites occurs on a faster timescale than 
vibrational relaxation, resulting in
energetic distributions that are intrinsically nonequilibrium.

\section{Acknowledgments}

AN is supported by the Israel Science Foundation, the US-Israel Bi-national Science Foundation and the University of Pennsylvania. 
AN thanks Prof. Michael Galperin for useful discussions.

\appendix

\section{\label{sec:appendix}Heat Current Derivation From Single-Electron Expectation Values}
To derive the expectation value for the heat transferred by an electron when it moves between environments,
we consider a single electron whose charge is localized on the molecular donor,
and the probability $P({\epsilon_0})$ for it to enter
the metal at energy $\epsilon_0$. 
This probability depends on multiple independent factors: (a) the probability $p_0$
that the solvent environment is in a configuration in which the
electron can be transferred into the metal at energy $\epsilon_0$,
(b) the probability $\big[1-f(\beta_\text{M},\epsilon_0)\big]$ that there is a vacancy in the metal
at energy $\epsilon_0$,
(c) the density of states in the metal $\rho_\text{M}(\epsilon_0)$ at energy $\epsilon_0$,
and (d) the probability that the transition between states occurs,
which can be calculated using the 
Landau-Zener expression \cite{Nitzan2006chemical}
and thus we denote this probability $P_\text{LZ}(\epsilon_0)$.
In the nonadiabatic limit, the general relationship $P_\text{LZ}(\epsilon) \rho_\text{M}(\epsilon) = \mathcal{T}\,\Gamma(\epsilon)$
holds,
where $\mathcal{T}$ is a constant that does not 
play a role in the calculations that follow, but we include it for completeness.

Because of the occupancy characteristics of each energy level in the
metal, the specific electron we consider
can make many attempts to enter the metal, at many different energies,
before the transfer event occurs.
We denote the probability of a successful attempt (meaning ET occurs) for transfer into the 
metal with energy $\epsilon_0$ as
$P_S(\epsilon_0) = \mathcal{T}\, \Gamma(\epsilon_0) \big[1-f(\beta_\text{M},\epsilon_0)\big] p_0$
where $p_0$ is the Boltzmann weight 
of energy level that leads to the electron entering the metal at energy $\epsilon_0$,
which depends on the temperature of the solvent environment of the molecular species.
However, due to the quasicontinuum of energy levels in the electronic manifold of the metal and the occupancy/vacancy
probabilities for each of these levels, there are many other possible outcomes for each attempt,
and each outcome must be accounted for to derive the amount of heat transferred.

The complexity of this network of events can be simplified by grouping them 
the into three possible outcomes for each attempt:
either the electron enters the metal at energy $\epsilon_0$ with probability $P_S(\epsilon_0)$,
the electron enters the metal at an energy that is not $\epsilon_0$, which we denote 
$\overline{P}_S(\epsilon_0)= \sum_k  \mathcal{T}\, \Gamma(\epsilon_k) \big[1-f(\beta_\text{M},\epsilon_k)\big] p_k - P_S(\epsilon_0)$,
or the electron attempts to transfer into the metal at any energy,
but there is no vacancy at the respective energy. 
We term the latter as an unsuccessful attempt 
and denote the probability for this outcome as $P_U$.
From conservation of probability for each attempt, $P_S(\epsilon_0) + \overline{P}_S(\epsilon_0)+ P_U = 1$, we find that
$P_U = 1 - \sum_k \mathcal{T}\, \Gamma(\epsilon_k)  \big[1-f(\beta_\text{M},\epsilon_k)\big] p_k$.
If the first electron transfer attempt is unsuccessful, 
the electron will eventually make another transfer attempt due to thermal fluctuations.
If the second attempt to transfer is also unsuccessful,
the electron will make a third attempt, 
and this process is repeated \textit{ad infinitum}.
If the first attempt is unsuccessful with probability $P_U$,
the probability for success on the second attempt given that the first was unsuccessful is
$P_U P_S(\epsilon_0)$, and the probability for success on a third attempt given that it is preceded by two previous
unsuccessful attempts is $P_U^2 P_S(\epsilon_0)$.
Taking the sum of all possible event sequences that lead to the 
electron entering the metal at energy $\epsilon_0$ gives
$P(\epsilon_0) = P_S(\epsilon_0)(1+P_U+P_U^2 + \cdots)$,
a geometric series.
The sum of this series can be expressed as $P(\epsilon_0) = P_S(\epsilon_0)/(1-P_U)$
which gives
$P(\epsilon_0) =  \Gamma(\epsilon_0) \big[1-f(\beta_\text{M},\epsilon_0)\big] p_0/\sum_k  \Gamma(\epsilon_k) \big[1-f(\beta_\text{M},\epsilon_k)\big] p_k$
after substitution for $P_U$.

This analysis can be performed for each energy level $\epsilon_j$ leading to the general expression
$P(\epsilon_j) = \Gamma(\epsilon_j) \big[1-f(\beta_\text{M},\epsilon_j)\big] p_j/\sum_k \Gamma(\epsilon_k) \big[1-f(\beta_\text{M},\epsilon_k)\big] p_k$.
Note that this derivation is also valid for the $b \to a$ transition in which the electron moves from metal to molecule,
provided that the probability of vacancy given by $\big[1-f(\beta_\text{M},\epsilon)\big]$ is replaced 
by the corresponding occupancy probability $f(\beta_\text{M},\epsilon)$ in each expression.

Using this event analysis to evaluate the probability of 
all possible transitions from the molecule into the metal,
and writing the sums in the $P(\epsilon_j)$ expression as integrals,
we find that
the Fermi-weighted configuration integral 
%(which arises from all combinations of successful and unsuccessful ET attempts described previously) 
%over possible activation energies 
for the $a \to b$ transition
(which corresponds to the denominator in the expression for $P(\epsilon_j)$ given above) is
\begin{equation}
\label{eq:parta}
\begin{aligned}
Z_{a \to b} &=  \mathcal{T}\int_{\mathbb{R}} \Gamma(\epsilon) \big[1-f(\beta_\text{M},\epsilon)\big] \\
& \qquad \times \exp\left[-\beta_\text{S} \frac{(-\Delta E_{ab} +\epsilon+E_\text{R})^2}{4 E_\text{R}}\right]\,d\epsilon. \\
\end{aligned}
\end{equation}
The expectation value of the heat supplied by the environment of the molecular species during the ascent (denoted by $\uparrow$) to the transition 
state on the $E_a$ surface is
\begin{equation}
\label{eq:moltometalheatfermiup}
\begin{aligned}
&\Big\langle \mathcal{Q}_{\text{S}}^{(a \to b)} \Big\rangle_\uparrow = -\frac{\mathcal{T}}{Z_{a \to b}} \int_{\mathbb{R}} \frac{(\Delta E_{ba} +\epsilon+E_\text{R})^2}{4 E_\text{R}} \Gamma(\epsilon)\\
& \times  \big[1-f(\beta_\text{M},\epsilon)\big] \exp\left[-\beta_\text{S} \frac{(-\Delta E_{ab} +\epsilon+E_\text{R})^2}{4 E_\text{R}}\right]\,d\epsilon,
\end{aligned}
\end{equation}
and for the descent (denoted by $\downarrow$) on the $E_b$ surface:
\begin{equation}
\label{eq:moltometalheatfermidown}
\begin{aligned}
&\Big\langle \mathcal{Q}_{\text{S}}^{(a \to b)} \Big\rangle_\downarrow = \frac{\mathcal{T}}{Z_{a \to b}} \int_{\mathbb{R}} \frac{(-\Delta E_{ba} -\epsilon+E_\text{R})^2}{4 E_\text{R}}  \Gamma(\epsilon)  \\
& \times  \big[1-f(\beta_\text{M},\epsilon)\big]\exp\left[-\beta_\text{S} \frac{(-\Delta E_{ab} +\epsilon+E_\text{R})^2}{4 E_\text{R}}\right]\,d\epsilon.
\end{aligned}
\end{equation}
The heat that flows into the metal during the $a \to b$ ET process is
\begin{equation}
\label{eq:moltometalheatfermimetal}
\begin{aligned}
&\Big\langle \mathcal{Q}_{\text{M}}^{(a \to b)} \Big\rangle = \frac{\mathcal{T}}{Z_{a \to b}} \int_{\mathbb{R}} (\epsilon-\mu) \Gamma(\epsilon) \big[1-f(\beta_\text{M},\epsilon)\big]  \\
& \qquad \times \exp\left[-\beta_\text{S} \frac{(-\Delta E_{ab} +\epsilon+E_\text{R})^2}{4 E_\text{R}}\right]\,d\epsilon.
\end{aligned}
\end{equation}
Note that if the Fermi factor was
not included in $Z_{a \to b}$,
Eqs.~(\ref{eq:moltometalheatfermiup})-(\ref{eq:moltometalheatfermimetal}) would give the 
respective expectation value per transition \textit{attempt};
with its inclusion these equations give the probability per transition \textit{event}.
The heat transferred to the solvent environment over the $a \to b$ transition is
\begin{equation}
\Big\langle \mathcal{Q}_\text{S}^{(a \to b)} \Big\rangle = \Big\langle \mathcal{Q}_\text{S}^{(a \to b)} \Big\rangle_\uparrow + \Big\langle \mathcal{Q}_\text{S}^{(a \to b)} \Big\rangle_\downarrow.
\end{equation}
The total free energy change by the molecular system and the metal is $\Delta E_{ba}+\mu$.
Correspondingly, by conservation of energy we expect that the environments must change by $-\Delta E_{ba}-\mu$.
We have verified, numerically, over a variety of parameter values, that the sum of the energy change during each leg of the
$a \to b$ transition gives
\begin{equation}
\label{eq:moltometalheatsumfermi}
\Big\langle \mathcal{Q}_\text{S}^{(a \to b)} \Big\rangle +\Big\langle \mathcal{Q}_\text{M}^{(a \to b)} \Big\rangle =  -\Delta E_{ba}-\mu,
\end{equation}
and thus that the expectation value expressions conserve energy.

For the $b \to a$ transition,
constructing the Fermi-weighted configuration integral yields
\begin{equation}
\label{eq:partb}
\begin{aligned}
Z_{b \to a} &= \mathcal{T}\int_{\mathbb{R}} \Gamma(\epsilon)f(\beta_\text{M},\epsilon)\\
& \qquad \times \exp\left[-\beta_\text{S} \frac{(\Delta E_{ab} -\epsilon+E_\text{R})^2}{4 E_\text{R}}\right]\,d\epsilon. \\
\end{aligned}
\end{equation}
The expectation value of the heat supplied by the environment of the molecular species during the ascent to the transition 
state on the $E_b$ surface is
\begin{equation}
\label{eq:moltometalheatfermiupba}
\begin{aligned}
&\Big\langle \mathcal{Q}_{\text{S}}^{(b \to a)} \Big\rangle_\uparrow = -\frac{\mathcal{T}}{Z_{b \to a}} \int_{\mathbb{R}} \frac{(-\Delta E_{ba} -\epsilon+E_\text{R})^2}{4 E_\text{R}} \Gamma(\epsilon) \\
& \times  f(\beta_\text{M},\epsilon)\exp\left[-\beta_\text{S} \frac{(\Delta E_{ab} -\epsilon+E_\text{R})^2}{4 E_\text{R}}\right]\,d\epsilon,
\end{aligned}
\end{equation}
and for the descent to equilibrium on the $E_a$ surface:
\begin{equation}
\label{eq:moltometalheatfermidownba}
\begin{aligned}
&\Big\langle \mathcal{Q}_{\text{S}}^{(b \to a)} \Big\rangle_\downarrow = \frac{\mathcal{T}}{Z_{b \to a}} \int_{\mathbb{R}} \frac{(\Delta E_{ba} +\epsilon+E_\text{R})^2}{4 E_\text{R}}  \Gamma(\epsilon) \\
& \times  f(\beta_\text{M},\epsilon) \exp\left[-\beta_\text{S} \frac{(\Delta E_{ab} -\epsilon+E_\text{R})^2}{4 E_\text{R}}\right]\,d\epsilon.
\end{aligned}
\end{equation}
The heat supplied by the metal is
\begin{equation}
\label{eq:moltometalheatfermimetalba}
\begin{aligned}
&\Big\langle \mathcal{Q}_{\text{M}}^{(b \to a)} \Big\rangle = \frac{\mathcal{T}}{Z_{b \to a}} \int_{\mathbb{R}} (\mu - \epsilon) \Gamma(\epsilon) f(\beta_\text{M},\epsilon)  \\
& \qquad \times \exp\left[-\beta_\text{S} \frac{(\Delta E_{ab} -\epsilon+E_\text{R})^2}{4 E_\text{R}}\right]\,d\epsilon.
\end{aligned}
\end{equation}
The heat transferred to the solvent environment over the $b \to a$ transition is
\begin{equation}
\Big\langle \mathcal{Q}_\text{S}^{(b \to a)} \Big\rangle = \Big\langle \mathcal{Q}_\text{S}^{(b \to a)} \Big\rangle_\uparrow + \Big\langle \mathcal{Q}_\text{S}^{(b \to a)} \Big\rangle_\downarrow.
\end{equation}
The sum of the free energy change by the molecular system and the metal is $-\Delta E_{ba}-\mu$,
and thus during this transition the environments must change by $\Delta E_{ba}+\mu$.
To confirm that our expectation value expressions conserve energy,
we take the sum of each process in the $b \to a$ transition (using numerically evaluation of the integrals).
In all studied cases we have found that
\begin{equation}
\label{eq:moltometalheatsumfermiba}
\Big\langle \mathcal{Q}_\text{S}^{(b \to a)} \Big\rangle +\Big\langle \mathcal{Q}_\text{M}^{(b \to a)} \Big\rangle =  \Delta E_{ba}+\mu,
\end{equation}
as expected.

The heat currents into the molecular environment and the metal are
\begin{equation}
\begin{aligned}
\label{eq:heatcurrent}
\dot{\mathcal{Q}}_\text{S} &= k_{a \to b} \mathcal{P}_a\Big\langle \mathcal{Q}_\text{S}^{(a \to b)} \Big\rangle+ k_{b \to a} \mathcal{P}_b \Big\langle \mathcal{Q}_\text{S}^{(b \to a)} \Big\rangle, \\[1ex]
\dot{\mathcal{Q}}_\text{M} &=  k_{a \to b} \mathcal{P}_a \Big\langle \mathcal{Q}_\text{M}^{(a \to b)} \Big\rangle+ k_{b \to a} \mathcal{P}_b \Big\langle \mathcal{Q}_\text{M}^{(b \to a)} \Big\rangle,
\end{aligned}
\end{equation}
respectively.
As in the case of homogeneous bithermal ET between molecules described in Ref.~\citenum{craven16c},
at steady-state, $k_{a \to b} \mathcal{P}^{(\text{ss})}_a = k_{b \to a} \mathcal{P}^{(\text{ss})}_b = \mathcal{J}_\text{ss}$,
and in this limit  the heat currents are
\begin{equation}
\begin{aligned}
\label{eq:heatcurrentss}
\dot{\mathcal{Q}}_\text{S} &= \mathcal{J}_\text{ss} \left(\Big\langle \mathcal{Q}_\text{S}^{(a \to b)} \Big\rangle+ \Big\langle \mathcal{Q}_\text{S}^{(b \to a)} \Big\rangle\right), \\[1ex]
\dot{\mathcal{Q}}_\text{M} &= \mathcal{J}_\text{ss} \left(\Big\langle \mathcal{Q}_\text{M}^{(a \to b)} \Big\rangle+ \Big\langle \mathcal{Q}_\text{M}^{(b \to a)} \Big\rangle\right),
\end{aligned}
\end{equation}
which agree with those derived in Eq.~(\ref{eq:heatcurrentAbe}).

%%%%%%%%%%%%%%%%%%%%%%
\bibliography{j,electron-transfer,craven,c2}

\begin{thebibliography}{80}
\bibitem[{not()}]{note2}
\bibinfo{note}{It is important to note that these results depend on the
  protocol used to define $\Delta T$ and $\Phi$. We will expand on this issue
  in a future paper.}
\end{thebibliography}
%\bibliography{c2_combined}
\end{document}